\documentclass[aps,prd,reprint,preprintnumbers,showpacs,floatfix,nofootinbib,superscript address]{revtex4-2}
\usepackage[utf8]{inputenc}
\usepackage{scalerel,stackengine}
\usepackage{titlesec}
\usepackage{parskip}
\usepackage{amssymb}
\usepackage{stix}
\usepackage{hhline}
\usepackage{amsmath}
\usepackage{mathtools}
\usepackage[dvipsnames]{xcolor}
\usepackage{xspace}
\usepackage{multirow,tabularx}
\usepackage{siunitx}
\usepackage{multirow}
\usepackage{graphicx}
\usepackage{xstring}
\usepackage{etoolbox}
\usepackage{notoccite}
\usepackage{natbib}
\usepackage{mathrsfs}
\usepackage{lineno}
\usepackage{tensor}
\usepackage{accents}
\usepackage{tikz}
\usepackage{bbm}
\usepackage[normalem]{ulem}

\allowdisplaybreaks

\newcommand{\cev}[1]{\reflectbox{\ensuremath{\vec{\reflectbox{\ensuremath{#1}}}}}}

\newcommand\lowerdag{{\mkern3mu\raise-1.3ex\hbox{$\scriptstyle\dag$}}}
\newcommand\lowerbar[1]{\mathstrut\mkern1.5mu#1\mkern-13mu\raise2.2ex
  \hbox{$\scriptscriptstyle-$}}

\newrobustcmd{\pea}[1]{%
	\emph{#1}\textbf{\ \ \ ---}
}
\titleformat{\paragraph}[runin]{\normalfont\normalsize\bfseries}{\emph\theparagraph}{1em}{\pea}

\newrobustcmd{\ovl}[1]{%
\overline{#1}%
}
\newrobustcmd{\lie}[1]{%
\hat{#1}%
}

\newrobustcmd{\hPerpZeroP}[1]{%
	{\tensor[^{0^+}]{\theta}{^{\perp}#1}}
}
\newrobustcmd{\hParaZeroP}[1]{%
	{\tensor[^{0^+}]{\theta}{^{\parallel}#1}}
}
\newrobustcmd{\hPerpOneM}[1]{%
	{\tensor[^{1^-}]{\theta}{^{\perp}#1}}
}
\newrobustcmd{\hParaOneM}[1]{%
	{\tensor[^{1^-}]{\theta}{^{\parallel}#1}}
}
\newrobustcmd{\hParaOneP}[1]{%
	{\tensor[^{1^+}]{\theta}{^{\parallel}#1}}
}
\newrobustcmd{\hParaTwoP}[1]{%
	{\tensor[^{2^+}]{\theta}{^{\parallel}#1}}
}

\newrobustcmd{\gpert}[1]{%
	\tensor{\varphi}{#1}
}
\newrobustcmd{\hpert}[1]{%
	\tensor{\theta}{#1}
}
\newrobustcmd{\hfield}[1]{%
	\tensor{h}{#1}
}
\newrobustcmd{\Hfield}[1]{%
	\tensor{H}{#1}
}
\newrobustcmd{\Hboldfield}[1]{%
	\tensor{\mathbf{H}}{#1}
}
\newrobustcmd{\Structure}[1]{%
	\tensor{f}{#1}
}
\newrobustcmd{\Gg}{%
	g_{\text{g}}	
}

\newrobustcmd{\g}[1]{%
	\tensor{g}{#1}%
}
\newrobustcmd{\rcCon}[1]{%
	\tensor*{\Gamma}{#1}%
}
\newrobustcmd{\rCon}[1]{%
	\tensor*{\mathring{\Gamma}}{#1}%
}
\newrobustcmd{\PD}[1]{%
	\tensor{\partial}{#1}%
}
\newrobustcmd{\rcD}[1]{%
	\tensor{\nabla}{#1}%
}
\newrobustcmd{\foli}[1]{%
	\tensor{n}{#1}%
}

\usepackage{hyperref}
\hypersetup{%
     colorlinks = true,%
     linkcolor = Blue,%
     citecolor = Blue,%
     filecolor = Blue,%
     urlcolor = Blue%
     }%
\usepackage[capitalize,nameinlink]{cleveref} 

\begin{document}

\title{Particle spectra of gravity based on internal symmetry of quantum fields}

\author{Will Barker}
\email{wb263@cam.ac.uk}
\affiliation{Astrophysics Group, Cavendish Laboratory, JJ Thomson Avenue, Cambridge CB3 0HE, UK}
\affiliation{Kavli Institute for Cosmology, Madingley Road, Cambridge CB3 0HA, UK}

\begin{abstract}
	We examine the weak-field, zero-coupling limit of Yang--Mills gravity as recently formulated by Partanen and Tulkki, viewed as a free quantum field theory. In this approximation the theory has a ghostly teleparallel vacuum. We suggest that bimetric, vacuum expectation value, or finite-coupling extensions should be investigated.
\end{abstract}

\maketitle

\paragraph*{The proposed model} In two recent pre-prints~\cite{Partanen:2023dkt,Partanen:2023sjn}, Partanen and Tulkki have proposed an $\mathrm{SU}(8)$ extension to the $\mathrm{U}(1)$ sector of the standard model of particle physics which aims to incorporate gravity through the methods of Yang and Mills. In a flat background, and setting $\hbar\equiv c\equiv 1$, their Lagrangian density is
\begin{align}
 \mathcal{L} &\equiv\frac{1}{4g_\mathrm{g}}\bar{\psi}(\cev{D}\bar{\mathbf{I}}_8\boldsymbol{\gamma}_\mathrm{B}^5\boldsymbol{\gamma}_\mathrm{B}^\nu\vec{\boldsymbol{\mathcal{D}}}_\nu\mathbf{I}_8\boldsymbol{\gamma}_\mathrm{F}-\bar{\boldsymbol{\gamma}}_\mathrm{F}\bar{\mathbf{I}}_8\boldsymbol{\gamma}_\mathrm{B}^5\boldsymbol{\gamma}_\mathrm{B}^\nu\vec{\boldsymbol{\mathcal{D}}}_\nu\mathbf{I}_8\vec{D})\psi\nonumber\\
 &\hspace{0.5cm}+\frac{i}{g_\mathrm{g}}\bar{\Psi}_\Re\mathbf{I}_8^\dag\boldsymbol{\gamma}_\mathrm{B}^5\boldsymbol{\gamma}_\mathrm{B}^\nu\lowerbar{\vec{\boldsymbol{\mathcal{D}}}}_\nu^\lowerdag\bar{\mathbf{I}}_8^\dag\Psi_\Re
 -m_\mathrm{e}\bar{\psi}\psi+\bar{\Psi}_\Re\Psi_\Re\nonumber\\
 &\hspace{0.5cm}
	+\frac{i}{2}\bar{\psi}\left(\bar{\boldsymbol{\gamma}}_\mathrm{F}\vec{D}-\cev{D}{\boldsymbol{\gamma}}_\mathrm{F}\right)\psi
	-\frac{1}{8\kappa}\Hfield{_{\lie{\rho}\mu\nu}}\Hfield{^{\lie{\rho}\mu\nu}}.
 \label{eq:Lagrangiancovariant}
\end{align}
In~\cref{eq:Lagrangiancovariant} an eight-spinor formalism refers to four $8\times8$ gamma matrices $\boldsymbol{\gamma}_\mathrm{B}^\mu$ and $\boldsymbol{\gamma}_\mathrm{B}^5\equiv i\boldsymbol{\gamma}_\mathrm{B}^0\boldsymbol{\gamma}_\mathrm{B}^1\boldsymbol{\gamma}_\mathrm{B}^2\boldsymbol{\gamma}_\mathrm{B}^3$, with specific components defined in~\cite{Partanen:2023sjn}, which satisfy the anticommutation relation $\{\boldsymbol{\gamma}_\mathrm{B}^\mu,\boldsymbol{\gamma}_\mathrm{B}^\nu\}\equiv 2\eta^{\mu\nu}\mathbf{I}_8$, for $\mathbf{I}_8$ the $8\times8$ identity and $\eta^{\mu\nu}$ the Minkowski metric tensor with the conventions $\eta^{\mu\nu}\equiv\mathrm{diag}(1,-1,-1,-1)$.
The spinors $\psi$ and $\Psi_\Re$ describe respectively the Dirac fermions and, additionally, the electromagnetic gauge bosons, where the latter are accommodated by the increased spinorial dimension. The eight-component derivative $\vec{D}$ is covariant with respect to the electromagnetic gauge transformation, whilst $\boldsymbol{\gamma}_\mathrm{F}$ is an eight-component version of the Dirac gamma matrices. These quantities account for all the non-gravitational `matter' in the theory. This matter sector is parameterised by the bare electron mass $m_\mathrm{e}$. The couplings between the matter and gravitational sectors are weighted by $1/\Gg$, and the gravitational kinetic operator is parameterised by the Einstein constant $\kappa$. Thus $\kappa$ and $\Gg$ are both gravitational couplings in this model. The $\Gg$ coupling also appears in the gravitational covariant derivative
\begin{equation}
	\vec{\boldsymbol{\mathcal{D}}}_\nu\equiv\mathbf{I}_8\vec{\partial}_\nu-ig_\mathrm{g}\mathbf{h}_\nu,\quad
	\lowerbar{\vec{\boldsymbol{\mathcal{D}}}}_\nu^\lowerdag\equiv\mathbf{I}_8\vec{\partial}_\nu-ig_\mathrm{g}\bar{\mathbf{h}}_\nu,
 \label{eq:covariantderivative}
\end{equation}
where $\mathbf{h}_\nu\equiv h_{\lie{\mu}\nu}\mathbf{t}^{\lie{\mu}}$ for $\hfield{_{\lie{\mu}\nu}}$ a dimensionless and a priori \emph{asymmetric} gauge field, and $\mathbf{t}^{\lie{\mu}}$ obeys the anticommutation relation $\{\mathbf{t}^{\lie{\mu}},\mathbf{t}^{\lie{\nu}}\}\equiv 2\delta^{\lie{\mu}\lie{\nu}}\mathbf{I}_8$, for $\delta^{\lie{\mu}\lie{\nu}}$ the Kronecker delta. Note that $\hat{\mu}$ and $\hat{\nu}$ are Lie indices, and this notation differs from~\cite{Partanen:2023dkt,Partanen:2023sjn}. Using~\cref{eq:covariantderivative} the gravitational kinetic operator is formed from $[\vec{\boldsymbol{\mathcal{D}}}_\mu,\vec{\boldsymbol{\mathcal{D}}}_\nu]\equiv-ig_\mathrm{g}\mathbf{H}_{\mu\nu}$, i.e. a non-Abelian field strength $\mathbf{H}_{\mu\nu}$ which expands to
\begin{subequations}
\begin{align}
	\mathbf{H}_{\mu\nu}&\equiv\partial_\mu\mathbf{h}_\nu-\partial_\nu\mathbf{h}_\mu-ig_\mathrm{g}[\mathbf{h}_\mu,\mathbf{h}_\nu]\equiv H_{\lie{\rho}\mu\nu}\mathbf{t}^{\lie{\rho}},\label{PreStrengthDefinition}\\
	\Hfield{_{\lie{\rho}\mu\nu}}&\equiv \PD{_\mu}\hfield{_{\lie{\rho}\nu}}-\PD{_\nu}\hfield{_{\lie{\rho}\mu}}+\Gg\Structure{_{\lie{\rho}}^{\lie{\sigma}\lie{\lambda}}}\hfield{_{\lie{\sigma}\mu}}\hfield{_{\lie{\lambda}\nu}},\label{StrengthDefinition}
\end{align}
\end{subequations}
where $\tensor{f}{^{\lie{\rho}\lie{\mu}\lie{\nu}}}\equiv 2\varepsilon^{0\lie{\rho}\lie{\mu}\lie{\nu}}$ for $\varepsilon^{\lie{\alpha}\lie{\beta}\lie{\gamma}\lie{\delta}}$ the Levi--Civita symbol in four spacetime dimensions. Note that instead of the Yang--Mills kinetic operator $\mathrm{Tr}(\mathbf{H}_{\mu\nu}\mathbf{H}^{\mu\nu})\equiv 8\tensor{\delta}{^{\lie{\rho}\lie{\sigma}}}\Hfield{_{\lie{\rho}\mu\nu}}\Hfield{_{\lie{\sigma}}^{\mu\nu}}$, the proposed operator in~\cref{eq:Lagrangiancovariant} contracts with the \emph{Minkowski} metric $\tensor{\eta}{^{\lie{\rho}\lie{\sigma}}}$, and this is put forwards as the correct metric with which to raise and lower all the Lie indices.
\paragraph*{Emergence of gravity} We can use~\cref{PreStrengthDefinition,StrengthDefinition} to effectively partition~\cref{eq:Lagrangiancovariant} as follows
\begin{equation}\label{GravityDefinition}
	\mathcal{L}\equiv\mathcal{L}_{0}+\mathcal{L}_{\text{g}},\quad
	\mathcal{L}_{\text{g}}\equiv-\frac{1}{8\kappa}\Hfield{_{\lie{\rho}\mu\nu}}\Hfield{^{\lie{\rho}\mu\nu}}.
\end{equation}
The matter Lagrangian density $\mathcal{L}_{0}$ reduces to Dirac's electrodynamics in the $\hfield{_{\lie{\mu}\nu}}\mapsto 0$ limit~\cite{Partanen:2023dkt,Partanen:2023sjn}. In the perturbative context, an identification with linear gravity is made in the ${\Gg\mapsto 0}$ limit of~\cref{StrengthDefinition}. 
The combination $\big(\tensor{\eta}{^{\lie{\sigma}}_{\mu}}+\hfield{^{\lie{\sigma}}_{\mu}}\big)\left(\tensor{\eta}{_{\lie{\sigma}\nu}}+\hfield{_{\lie{\sigma}\nu}}\right)$, with two lowered spacetime indices, is regarded as the gravitationally curved metric on the $\delta\mathcal{L}_{\text{g}}/\delta\hfield{_{\lie{\mu}\nu}}\approx 0$ shell. Coupling to a conserved electromagnetic stress-energy tensor can be provided by re-introducing $\mathcal{L}_{0}$, whilst the theory at finite $\Gg$ presumably has nonlinear modifications.

\paragraph*{Things we can easily check} We are anxious to explore the basic health of~\cref{GravityDefinition} in the non-interacting limit around motivated backgrounds. The most important and tractable background is clearly that of Minkowski, the setting in which the model is formulated. Until very recently, the task of obtaining the spectrum of propagating quantum particle states --- even for \emph{free} tensorial field theories --- was a potentially arduous one. However this information can now be rapidly determined with the theory-agnostic PSALTer software~\cite{PSALTer}, which enumerates all the propagator poles, their masses and residues, and also the total no-ghost and no-tachyon conditions which are implied among the Lagrangian couplings. We will now probe the quantum spectra implied by the gravitational sector of~\cite{Partanen:2023dkt,Partanen:2023sjn}, under various interpretations of the model.

\begin{figure*}[t!]
\includegraphics[scale=.43]{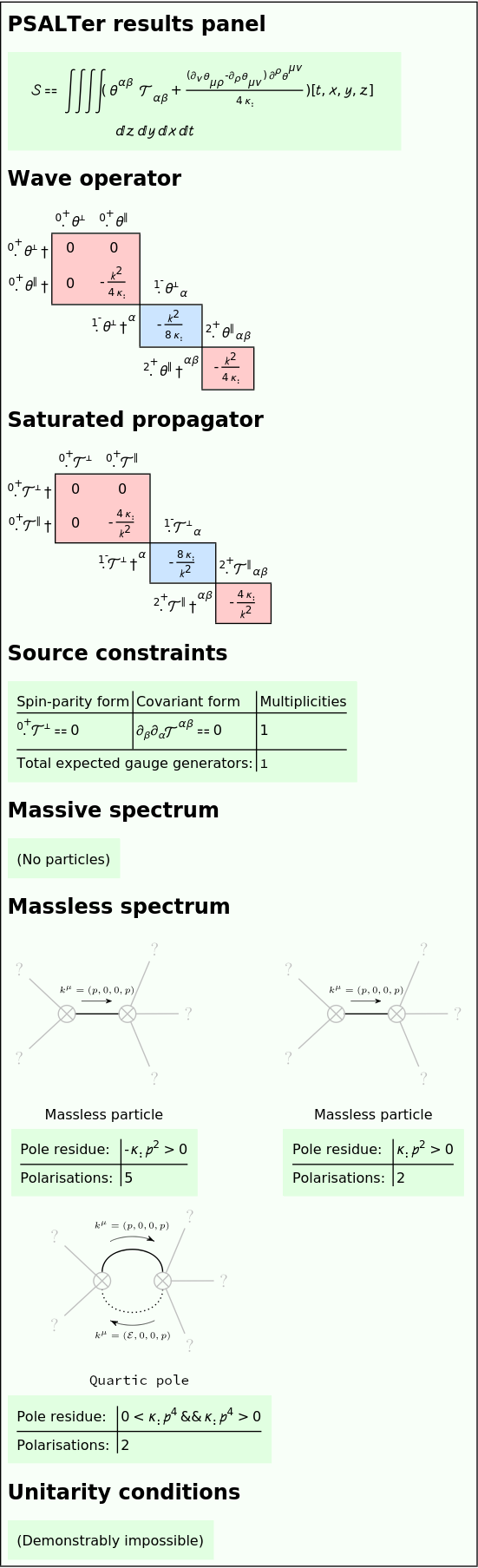} \hfill
\includegraphics[scale=.43]{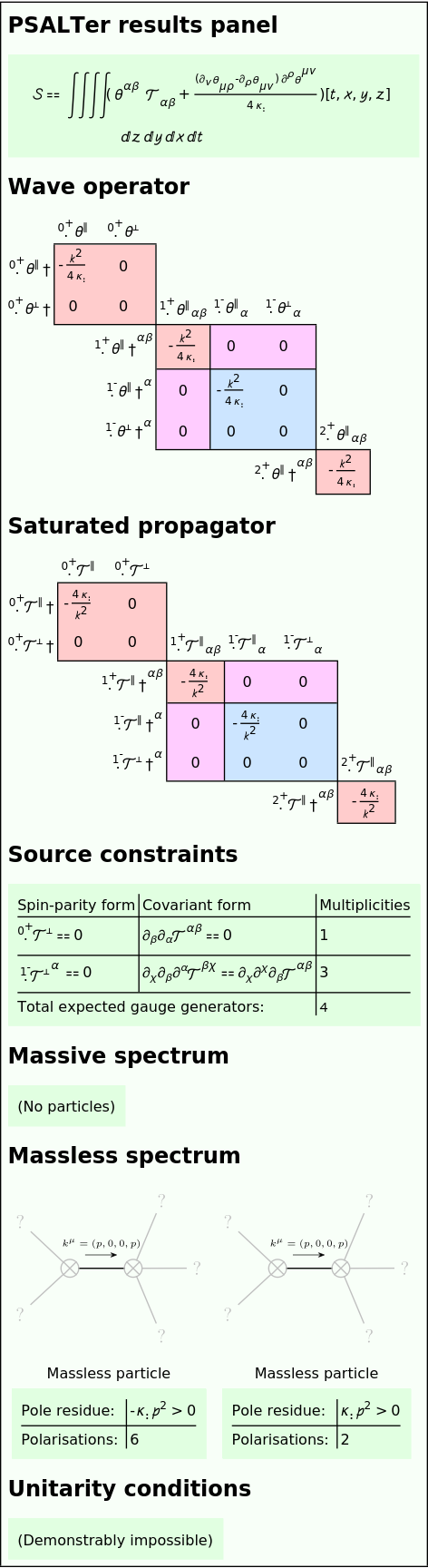} \hfill
\includegraphics[scale=.25]{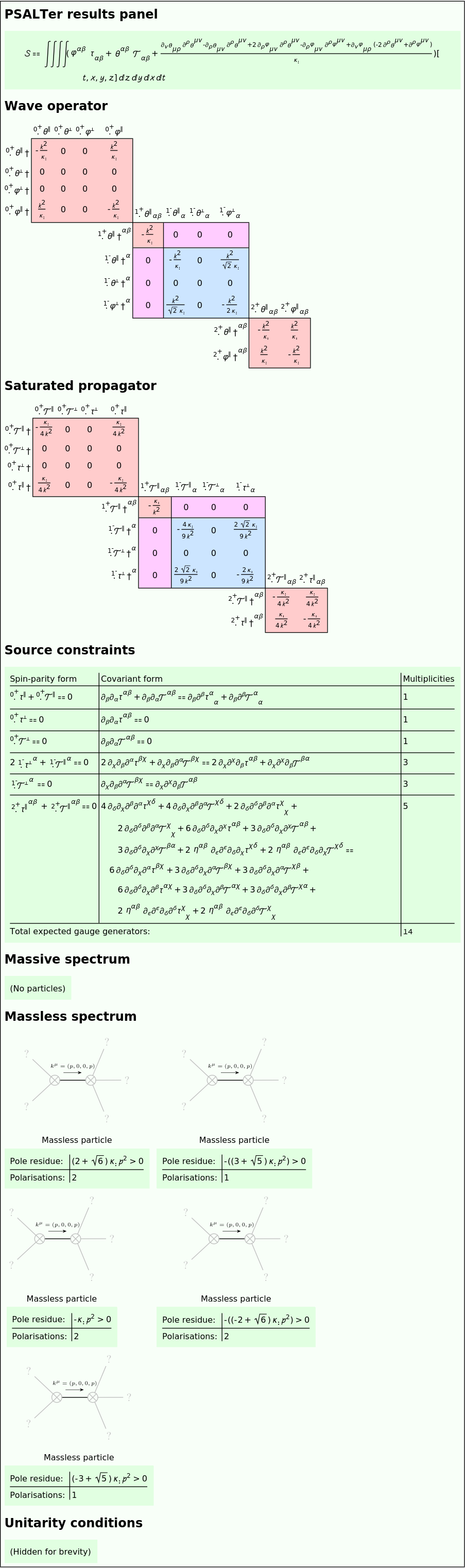}
	\caption{Particle spectra obtained by the PSALTer software, where $p\equiv\tensor{k}{^0}$ is the energy-momentum in the null limit $\tensor{k}{_\mu}\tensor{k}{^\mu}\mapsto 0$, and all the other symbols are as defined in the text. Left: the \emph{symmetric} field with $J^P$ sectors as in~\cref{SymmetricDecomposition} and the Lagrangian density in~\cref{LinearLagrangian}. Centre: the \emph{asymmetric} field with $J^P$ sectors as in~\cref{AsymmetricDecomposition} and the same Lagrangian density. Right: the \emph{bimetric} interpretation with the Lagrangian density in~\cref{BimetricLagrangian}. All cases would simultaneously require $\kappa>0$ and $\kappa<0$ in order to avoid multiple ghost polarisations, and are therefore not consistent as perturbative limits to gravity. As shown in~\cite{Supplement} the origin of some of these problems is in higher-order poles.}
\label{PSALTerOutput}
\end{figure*}

\paragraph*{Symmetric view} Since the Lie indices are supposed to be raised and lowered by the Minkowski metric, we will henceforth drop their special decoration. 
It is nowhere assumed in~\cite{Partanen:2023dkt,Partanen:2023sjn} that $\hfield{_{\mu\nu}}\equiv\hfield{_{(\mu\nu)}}$. However, the space of non-interacting, symmetric rank-two field theories has been well studied~\cite{VanNieuwenhuizen:1973fi}, making this a comfortable starting point. We also note that whilst~\cite{Partanen:2023dkt} takes the gauge field to be inherently perturbative, there is an obvious shift symmetry in the $\Gg\mapsto 0$ limit that allows us to expand around a Minkowskian vacuum expectation value (VEV) without changing any of our conclusions.
Near this VEV, we make the \emph{exact} re-parameterisation $\hfield{_{\mu\nu}}\equiv\tensor{\eta}{_{\mu\nu}}+\hpert{_{\mu\nu}}$, which has the same (exact) form with raised indices $\hfield{^{\mu\nu}}\equiv\tensor{\eta}{^{\mu\nu}}+\hpert{^{\mu\nu}}$, since only the flat metric is involved. All field perturbations will carry an implicit perturbative parameter $\epsilon$, so we can say e.g. $\hpert{^{\mu\nu}}\equiv\mathcal{O}(\epsilon)$. Let $\foli{_\mu}\equiv\tensor{k}{_{\mu}}/k$ be a unit-timelike vector for $k^2\equiv\tensor{k}{^\mu}\tensor{k}{_\mu}$, where $\tensor{k}{_\mu}$ is the (massive) particle four-momentum. With respect to this frame, the following spin-parity (we use the notation $J^P$ for spin $J$ and parity $P$) states are contained within the dynamical field
\begin{equation}\label{SymmetricDecomposition}
	\hpert{_{\mu\nu}}\equiv\foli{_\mu}\foli{_\nu}\hPerpZeroP{}+2\foli{_{(\mu}}\hPerpOneM{_{\ovl{\nu})}}+\hParaTwoP{_{\ovl{\mu\nu}}}+\frac{1}{3}\tensor{\eta}{_{\ovl{\mu\nu}}}\hParaZeroP{},
\end{equation}
where the over-bar denotes projection onto the Cauchy-slice, i.e. orthogonality of the index upon contraction with $\foli{_\mu}$, and the irreducible parts are defined in~\crefrange{ParaZeroP}{ParaTwoP}. Let the (symmetric, perturbative stress-energy) source ${\tensor{T}{^{\mu\nu}}\equiv\mathcal{O}(\epsilon)}$ be conjugate to $\hpert{_{\mu\nu}}$, having an equivalent decomposition to that in~\cref{SymmetricDecomposition} where we replace the symbol `$\theta$' with `$T$'. This source thus encodes all the gravitational couplings within the matter sector $\mathcal{L}_0$, whilst keeping the matter dynamics entirely anonymous. We perform a spectral analysis in this notation by plugging the free theory (in the $\Gg\mapsto 0$ limit) of~\cref{GravityDefinition}
\begin{equation}\label{LinearLagrangian}
	\mathcal{L}\equiv-\frac{1}{2\kappa}\PD{_{[\mu|}}\hpert{_{\rho|\nu]}}\PD{^{[\mu|}}\hpert{^{\rho|\nu]}}+\hpert{_{\mu\nu}}\tensor{T}{^{\mu\nu}}+\mathcal{O}\big(\epsilon^3\big),
\end{equation}
into the PSALTer software, and the result is shown in~\cref{PSALTerOutput}. The software correctly identifies in~\cref{LinearLagrangian} the Hessian symmetry $\hpert{_{\mu\nu}}\mapsto\hpert{_{\mu\nu}}+\PD{_\mu}\PD{_\nu}\zeta$ for some scalar generator $\zeta$, but no further symmetries. Since~\cref{LinearLagrangian} lacks any intrinsic scale under canonical normalisaiton, there are \emph{no} massive species. The massless species include the two polarisations (consistent with the Einstein graviton) whose unitarity naturally requires ${\kappa>0}$. However, accompanying this particle are \emph{five} unexpected polarisations whose combined unitarity requires ${\kappa<0}$. These are ghosts, so the symmetric model is non-viable.
\paragraph*{Asymmetric and teleparallel views} Now we relax the symmetry assumption on $\hfield{_{\mu\nu}}$ and its $\mathcal{O}(\epsilon)$ perturbation $\hpert{_{\mu\nu}}$, so that six further degrees of freedom (d.o.f) must be included in~\cref{SymmetricDecomposition}
\begin{equation}\label{AsymmetricDecomposition}
	\begin{aligned}
		\hpert{_{\mu\nu}}&\equiv\foli{_\mu}\foli{_\nu}\hPerpZeroP{}+\foli{_{\mu}}\hParaOneM{_{\ovl{\nu}}}+\foli{_{\nu}}\hPerpOneM{_{\ovl{\mu}}}
	\\
		&\ \ \ +\hParaOneP{_{\ovl{\mu\nu}}}+\hParaTwoP{_{\ovl{\mu\nu}}}+\frac{1}{3}\tensor{\eta}{_{\ovl{\mu\nu}}}\hParaZeroP{},
	\end{aligned}
\end{equation}
again these are all provided in~\crefrange{ParaZeroP}{ParaTwoP}.
This time in~\cref{PSALTerOutput} there is a full diffeomorphism (diff.) symmetry in~\cref{LinearLagrangian}, of the form $\hpert{_{\mu\nu}}\mapsto\hpert{_{\mu\nu}}+\PD{_\nu}\tensor{\xi}{_\mu}$ for some vector generator $\tensor{\xi}{^\mu}$. The presence of the diff. symmetry is suggestive of gravity and --- as with the symmetric $\hfield{_{\mu\nu}}$ case --- the Einstein graviton reappears. However this time it is accompanied by \emph{six} ghost polarisations. It is helpful to consider the following interpretation of~\cref{LinearLagrangian}. Let $\tensor{e}{^i_\mu}$ and $\tensor{e}{_i^\mu}$ be the co-tetrad and tetrad components associated with Roman Lorentz (i.e. anholonomic) indices, so that a curved background would have metric $\tensor{e}{^i_\mu}\tensor{e}{^j_\nu}\tensor{\eta}{_{ij}}\equiv\tensor{g}{_{\mu\nu}}$ and inverse $\tensor{e}{_i^\mu}\tensor{e}{_j^\nu}\tensor{\eta}{^{ij}}\equiv\tensor{g}{^{\mu\nu}}$ with $\tensor{e}{^i_\mu}\tensor{e}{_i^\nu}\equiv\tensor*{\delta}{_\mu^\nu}$ and $\tensor{e}{^i_\mu}\tensor{e}{_j^\mu}\equiv\tensor*{\delta}{_j^i}$ as kinematic restrictions. In the Weitzenböck gauge we may neglect the spin connection, and consider a teleparallel geometry in which the torsion tensor is $\tensor{\mathcal{T}}{^\lambda_{\mu\nu}}\equiv 2\tensor{e}{_k^\lambda}\PD{_{[\mu}}\tensor{e}{^k_{\nu]}}$. Clearly~\cref{LinearLagrangian} is also the free $\mathcal{O}\big(\epsilon^2\big)$ limit of the theory $\mathcal{L}_{\text{g}}\equiv-\frac{1}{8\kappa}\sqrt{-g}\tensor{\mathcal{T}}{^\lambda_{\mu\nu}}\tensor{\mathcal{T}}{_\lambda^{\mu\nu}}$ near the Kronecker (Minkowski) tetrad (metric) vacuum\footnote{Near this vacuum the Roman Lorentz and Greek holonomic indices are indistinguishable.} of this geometry, accompanied by some suitable matter sector: that theory also has diff. symmetry, but it is \emph{not} the uniquely healthy teleparallel theory~\cite{Ortin:2015hya,BeltranJimenez:2019nns,Blagojevic:2002du}\footnote{See also a more liberal space of viable theories considered in~\cite{Golovnev:2023ddv,Golovnev:2023uqb}.}.

\paragraph*{Bimetric interpretation} We can also consider a more familiar framework for introducing gravity, in which the background metric is \emph{also} made dynamical~\cite{Schmidt-May:2015vnx}. Note that this is not explicitly suggested in~\cite{Partanen:2023dkt,Partanen:2023sjn}, but it is very easy to try. Purely as a demonstration, we promote the Minkowski background spacetime metric, and also the Minkowski Lie index metric, to the same dynamical field $\tensor{g}{_{\mu\nu}}$, which is a priori independent of the gauge field. This time, both the dynamical metric and the gauge field will be expanded around a Minkowski VEV. Let the \emph{exact} metric perturbation be $\tensor{g}{_{\mu\nu}}\equiv\tensor{\eta}{_{\mu\nu}}+\gpert{_{\mu\nu}}$, so that $\tensor{g}{^{\mu\nu}}\equiv\tensor{\eta}{^{\mu\nu}}-\gpert{^{\mu\nu}}+\mathcal{O}\big(\epsilon^2\big)$ is an infinite series with the change in sign required by the kinematic restriction $\tensor{g}{_{\mu\nu}}\tensor{g}{^{\nu\lambda}}\equiv\tensor*{\delta}{^\lambda_\mu}$, and let the conjugate source to $\gpert{_{\mu\nu}}$ be $\tensor{\tau}{^{\mu\nu}}\equiv\mathcal{O}(\epsilon)$. 
The symmetry $\tensor{g}{_{\mu\nu}}\equiv\tensor{g}{_{(\mu\nu)}}$ is another kinematic restriction, and so the $J^P$ states of the metric perturbation field and its conjugate source will be precisely those in~\cref{SymmetricDecomposition}, replacing the symbol `$\theta$' with `$\varphi$' and `$\tau$'.
As above and in~\cite{Partanen:2023dkt,Partanen:2023sjn}, we will not assume any kinematic restriction on the exactly-defined, asymmetric $\tensor{h}{_{\mu\nu}}\equiv\tensor{\eta}{_{\mu\nu}}+\hpert{_{\mu\nu}}$, though the raised-index version $\tensor{h}{^{\mu\nu}}\equiv\tensor{\eta}{^{\mu\nu}}+\hpert{^{\mu\nu}}-\gpert{^{\mu\nu}}+\mathcal{O}\big(\epsilon^2\big)$ also now becomes an infinite series. With this prescription, the free bimetric theory (see e.g.~\cite{Blixt:2023qbg} for bimetric models including asymmetric or antisymmetric fields) is the $\mathcal{O}\big(\epsilon^2\big)$ part of the \emph{covariantised} action, so~\cref{LinearLagrangian} becomes
\begin{align}\label{BimetricLagrangian}
	\mathcal{L}&\equiv-\frac{1}{2\kappa}\sqrt{-g}\rcD{_{[\mu|}}\hfield{_{\rho|\nu]}}\rcD{^{[\mu|}}\hfield{^{\rho|\nu]}}+\mathcal{L}_0
	\nonumber
	\\
	&=-\frac{1}{2\kappa}\Big(\PD{_{[\mu|}}\gpert{_{\rho|\nu]}}-\PD{_{[\mu|}}\hpert{_{\rho|\nu]}}\Big)\Big(\PD{^{[\mu|}}\gpert{^{\rho|\nu]}}-\PD{^{[\mu|}}\hpert{^{\rho|\nu]}}\Big)
	\nonumber
	\\
	&\ \ \ \ \
	+\hpert{_{\mu\nu}}\tensor{T}{^{\mu\nu}}+\gpert{_{\mu\nu}}\tensor{\tau}{^{\mu\nu}}+\mathcal{O}\big(\epsilon^3\big).
\end{align}
Once again the spectrum of~\cref{BimetricLagrangian} is provided in~\cref{PSALTerOutput}. In this particular configuration, the theory is still not unitary. 

\paragraph*{Closing remarks} The roles of the background metric and the gauge field in the theory of~\cite{Partanen:2023dkt,Partanen:2023sjn} remain open to interpretation. We have probed the graviton spectrum under certain naïve and arbitrary assumptions, but we did not yet encounter the familiar Einsteinian limit. However, we did illustrate the ease with which spectra may be obtained, once the assumptions are specified.

Regarding further developments of~\cite{Partanen:2023dkt,Partanen:2023sjn} beyond just the simple steps presented here, we recommend the physical investigation of any bimetric, VEV, finite-$\Gg$ or interpretive extensions which may be put forward as part of a mathematically consistent formulation.

\begin{acknowledgments}

WB is grateful for useful correspondence with Mikko Partanen, and other useful discussions with Daniel Blixt, Mike Hobson, Anthony Lasenby, Carlo Marzo, Tobias Mistele, Syksy Räsänen, Sebastian Zell and Tom Złośnik.

This work was performed using resources provided by the Cambridge Service for Data Driven Discovery (CSD3) operated by the University of Cambridge Research Computing Service (\href{www.csd3.cam.ac.uk}{www.csd3.cam.ac.uk}), provided by Dell EMC and Intel using Tier-2 funding from the Engineering and Physical Sciences Research Council (capital grant EP/T022159/1), and DiRAC funding from the Science and Technology Facilities Council (\href{www.dirac.ac.uk}{www.dirac.ac.uk}).

This work was also performed using the Newton server, access to which was provisioned by Will Handley.

WB is grateful for the kind hospitality of Leiden University and the Lorentz Institute, and the support of Girton College, Cambridge.
\end{acknowledgments}

\bibliographystyle{apsrev4-1}
\bibliography{Manuscript}

\appendix

\section{Spin-parity decomposition} \label{Appendix}
The conventions used for the $J^P$ irreducible parts of the \emph{asymmetric} field in~\cref{AsymmetricDecomposition} are
\begin{subequations}
	\begin{align}
		\hParaZeroP{}
		&\equiv
		\hpert{^{\ovl{\mu}}_{\ovl{\mu}}},
		\label{ParaZeroP}
		\\
		\hPerpZeroP{}
		&\equiv
		\hpert{_{\perp\perp}},
		\\
		\hParaOneP{_{\ovl{\mu\nu}}}
		&\equiv
		\hpert{_{[\ovl{\mu\nu}]}},
		\\
		\hParaOneM{_{\ovl{\nu}}}
		&\equiv
		\hpert{^{\perp}_{\ovl{\nu}}},
		\\
		\hPerpOneM{_{\ovl{\mu}}}
		&\equiv
		\hpert{_{\ovl{\mu}}^{\perp}},
		\\
		\hParaTwoP{_{\ovl{\mu\nu}}}
		&\equiv
		\hpert{_{(\ovl{\mu\nu})}}-\frac{1}{3}\tensor{\eta}{_{\ovl{\mu\nu}}}\hParaZeroP{}
		,
		\label{ParaTwoP}
	\end{align}
\end{subequations}
where `$\perp$' denotes contraction of the index with $\foli{_\mu}$. These conventions are shared by the \emph{symmetric} field in~\cref{SymmetricDecomposition}, with the symmetric limit being $\hParaOneP{_{\ovl{\mu\nu}}}\mapsto 0$ and $\hParaOneM{_{\ovl{\nu}}}\mapsto \hPerpOneM{_{\ovl{\nu}}}$, i.e. a loss of six d.o.f.
\end{document}